\documentclass[%
reprint,
 amsmath,
 amssymb,
 noeprint,
 aps,
 prl,
 superscriptaddress,
floatfix,
]{revtex4-2}

\usepackage{graphicx}
\usepackage{dcolumn}
\usepackage{bm}

\newlength{\onecolfig}
\newlength{\twocolfig}
\usepackage{physics}
\usepackage{siunitx}
\usepackage{braket}

\makeatletter
\renewcommand*{\@fnsymbol}[1]{\ensuremath{\ifcase#1\or \dagger\or *\or \ddagger\or
    \mathsection\or \mathparagraph\or \|\or **\or \dagger\dagger
    \or \ddagger\ddagger \else\@ctrerr\fi}}
\makeatother

\newcommand{\hc}{\,\mathrm{h.c.}}
\newcommand{\G}{\,\mathrm{G}}

\newcommand{\kHz}{\,\mathrm{kHz}}
\newcommand{\MHz}{\,\mathrm{MHz}}
\newcommand{\GHz}{\,\mathrm{GHz}}
\newcommand{\THz}{\,\mathrm{THz}}
\newcommand{\um}{\,\mathrm{\mu m}}

\newcommand{\nm}{\,\mathrm{nm}}

\newcommand{\us}{\,\mathrm{\mu s}}
\newcommand{\ms}{\,\mathrm{ms}}

\newcommand{\mW}{\,\mathrm{mW}}

\newcommand{\ca}{\ensuremath{\mathrm{Ca}^+\,}}

\newcommand{\ct}{\ensuremath{^{43}\mathrm{Ca}^+\,}}
\newcommand{\sr}{\ensuremath{^{88}\mathrm{Sr}^+\,}}

\newcommand{\ddKet}{\ensuremath{\ket{\Downarrow \downarrow}}\,}

\newcommand{\upa}{\ensuremath{\uparrow}}
\newcommand{\dna}{\ensuremath{\downarrow}}

\newcommand{\dn}{\ensuremath{\ket{\downarrow}}\,}

\newcommand{\Dn}{\ensuremath{\ket{\Downarrow}}\,}

\newcommand{\psp}{\ensuremath{\sigma_+}}
\newcommand{\psm}{\ensuremath{\sigma_-}}
\newcommand{\psz}{\ensuremath{\sigma_z}}
\newcommand{\psx}{\ensuremath{\sigma_x}}
\newcommand{\psy}{\ensuremath{\sigma_y}}
\newcommand{\psphi}{\ensuremath{\sigma_\phi}}

\newcommand{\wz}{\ensuremath{\omega_{z}}}
\newcommand{\wo}{\ensuremath{\omega_{0}}}

\newcommand{\dlg}{\ensuremath{\delta_\mathrm{g}}}
\newcommand{\ad}{\ensuremath{a^{\dagger}}}
\newcommand{\zh}{\ensuremath{\hat{\bf z}}}

\newcommand{\Oj}{\ensuremath{\Omega_{j}}}

\newcommand{\phd}{\ensuremath{\phi_\mathrm{d}}}
\newcommand{\phs}{\ensuremath{\phi_\mathrm{s}}}
\newcommand{\php}{\ensuremath{\phi_\mathrm{+}}}
\newcommand{\phm}{\ensuremath{\phi_\mathrm{-}}}

\newcommand{\zg}{\ensuremath{\psz\otimes\psz\ }}
\newcommand{\phig}{\ensuremath{\psphi\otimes\psphi\ }}
\newcommand{\MS}{M\o lmer-S\o rensen\ }

\begin{document}

\preprint{APS/123-QED}

\title{Comparison of trapped-ion entangling gate mechanisms for mixed species}

\author{V.\,M. Sch\"afer}
\affiliation{Department of Physics, University of Oxford, Clarendon Laboratory, Parks Road, Oxford OX1 3PU, U.K.}
\affiliation{Max-Planck-Institut f\"ur Kernphysik, Saupfercheckweg 1, 69117 Heidelberg, Germany}
\author{A.\,C. Hughes}
\affiliation{Department of Physics, University of Oxford, Clarendon Laboratory, Parks Road, Oxford OX1 3PU, U.K.}
\author{O. Bazavan}
\affiliation{Department of Physics, University of Oxford, Clarendon Laboratory, Parks Road, Oxford OX1 3PU, U.K.}
\author{K. Thirumalai}
\affiliation{Department of Physics, University of Oxford, Clarendon Laboratory, Parks Road, Oxford OX1 3PU, U.K.}
\author{G. Pagano}
\affiliation{Department of Physics, University of Oxford, Clarendon Laboratory, Parks Road, Oxford OX1 3PU, U.K.}
\affiliation{Department of Physics and Astronomy and Smalley-Curl Institute, Rice University, Houston, TX 77005, USA}
\author{C.\,J. Ballance}
\affiliation{Department of Physics, University of Oxford, Clarendon Laboratory, Parks Road, Oxford OX1 3PU, U.K.}
\author{D.\,M. Lucas}
\affiliation{Department of Physics, University of Oxford, Clarendon Laboratory, Parks Road, Oxford OX1 3PU, U.K.}

\date{\today}

\begin{abstract}
Entangling gates are an essential capability of quantum computers.
There are different methods for implementing two-qubit gates, with respective advantages and disadvantages.
We investigate the experimentally relevant differences and commonalities of laser-based \zg light-shift and \phig \MS gates, highlighting the phases of experimental control fields and their long-term stabilities, in the specific case of mixed-species gates.
We implement these gates on qubits with very different magnetic field sensitivities, encoded in \ct and \sr\!, achieving fidelities of $99.8\%$ for the \zg and $99.6\%$ for the \phig gate.
\end{abstract}


\maketitle

Trapped ions are a cornerstone in atomic physics, with applications in quantum computing~\cite{cirac1995quantum, monroe1995demonstration, Wineland1998}, optical networking~\cite{monroe1995demonstration, stephenson2020high}, quantum simulation~\cite{ blatt2012quantum}, atomic clocks and precision measurements for tests of fundamental physics~\cite{schmidt2005spectroscopy, roseband2008frequency, wolf2016non, brewer2019quantum,Door2025,Filzinger2023}.
Different ion species have distinct properties that make them better suited for individual applications, such as transitions suitable for particular cooling schemes, with suppressed error sources for precision measurements and quantum computing, or fewer decay paths ideal for networking~\cite{ barrett2003sympathetic, home2009memory,nigmatullin2016ramil,inlek2017multispecies}. 
Co-trapping different ion species in the same trap greatly increases the scope of trapped-ion experiments, as the capabilities of both species can be leveraged.
It also allows for individually addressed logic or readout operations thanks to high spectral isolation between different species, or the indirect probing and manipulation of exotic species via quantum logic spectroscopy~\cite{schmidt2005spectroscopy,roseband2008frequency, Negnevitsky2018,Micke2020}.
To take full advantage of mixed-species experiments, it is necessary to coherently transfer superpositions of internal states of the two ions.
This can be achieved with two-qubit entangling gates~\cite{inlek2017multispecies, Ballance2015, Tan2015, Negnevitsky2018, Bruzewicz2019, Wan2019, hughes2020benchmarking}, which play a crucial role in mixed-species quantum information processing~\cite{drmota2023robust, drmota2024verifiable,main2024distributed} and can provide metrological gain~\cite{nichol2022elementary}.
While same-species entangling gates are widely used and achieve consistently improving error rates~\cite{Ballance2016, Gaebler2016, clark2021high, srinivas2021high, loschnauer2024scalable},
mixed-species gates pose more difficulties, as different masses reduce the motional coupling between the ions and the qubits typically have different magnetic field sensitivities~\cite{home2013quantum}.
In this paper we compare the two most prominent classes of laser-based trapped-ion entangling gates, \MS gates~\cite{sorensen1999quantum, sorensen2000entanglement, sackett2000experimental, Gaebler2016, erhard2019characterizing} and light-shift gates~\cite{milburn2000ion, Leibfried2003, Ballance2016}, with respect to their application to mixed-species crystals.
Both are geometric phase gates where the two ions' spin states are entangled via excitation of the collective state of motion of the ions.
We detail their sensitivities to different parameters, especially to different phases, and the consequences for the robustness and attainable fidelity for mixed-species entangling gates.
Finally, we perform an experimental comparison of the different techniques in the same system, a \ct $-$\sr ion crystal, and compare the fidelities achieved.

\section{Hamiltonians and bases}

The most commonly used trapped-ion entangling gates are the \MS (MS) and the light-shift (LS) gate.
Their first-order insensitivity to the temperature of the ion crystal makes them robust against an otherwise major source of error and facilitates the achievement of high fidelities.
As detailed by P.J.\,Lee et al.~\cite{Lee2005}, the two mechanisms are closely related and their effective Hamiltonians are equivalent apart from being in a rotated basis: the MS gate operates in the \phig basis, while the LS gate operates in the \zg basis.
However, this basis rotation leads to different sensitivities to different phases for the two mechanisms.
For mixed-species crystals, due to the lifting of degeneracies for several parameters, some of these sensitivities play an enhanced role.\\

The MS gate is driven by a bichromatic field composed of two tones detuned by the gate detuning $\delta_g$ from the red and blue sideband transitions $\pm\wz$ of the qubit transition at $\omega_0$~\cite{sorensen1999quantum, sorensen2000entanglement}.
The Hamiltonian in the interaction picture after making the Lamb-Dicke approximation $\eta \ll 1$, i.e. to first order in the Lamb-Dicke parameter $\eta$ is

\begin{align}
H_\mathrm{MS}(t)= \sum_{j=1,2}\sum_{k=+,-} & - \frac{\hbar \Omega_{k,j}}{2} \bigg[ \left(1+ i\eta_j\left[ae^{-i\wz t}+\ad e^{i\wz t}\right]\right) \nonumber\\
 & e^{i(\phi_{k,j}-\tilde{\delta}_k t)} \psp  +\hc \bigg]
\end{align}

\noindent where $j=1,2$ is the ion index, $k=+,-$ is the index of the frequency tone $\delta_k=\omega_0+\tilde{\delta}_k=\omega_0\pm(\omega_z + \delta_g)$ with detuning $\tilde{\delta}_k$, $\omega_z$ is the motional mode frequency, $a/\ad$ are the lowering/raising operators for this motional mode, $\eta_j$ is the Lamb-Dicke factor of the light coupling to the respective ion, $\psp,\psm$ are Pauli matrices, and $\phi_{k,j}$ is the phase of the individual driving tones at ion $j$.
After making the rotating wave approximation, for $\Omega_+=\Omega_-$ and neglecting the carrier term which does not couple to the motion, this simplifies to 

\begin{align}
H_\mathrm{MS}(t)= \sum_{j=1,2} & \frac{\hbar \Oj}{2} \eta_{j}  \sigma_{\frac{\pi}{2}-\phi_{s,j},j} \nonumber \\  
& \Big[  a  e^{  i(\dlg t -\phi_{d,j})}  + \ad e^{-i(\dlg t -\phi_{d,j})}   \Big]
\label{eqn:MS_ham}
\end{align}

where we used the sum and difference phases $\phs=\frac{1}{2}(\php+\phm)$ and $\phd=\frac{1}{2}(\php-\phm)$, and the rotated Pauli operator $\sigma_\phi=\cos(\phi)\psx+\sin(\phi)\psy$, with $\sigma_{\phi,j=1}=\sigma_\phi \otimes I$.

The LS gate is driven by two tones that are far off-resonant from any transitions, but whose frequency difference $\omega$ is close to a motional resonance of the ions.
The two beams interfere to form a travelling standing wave, which gives rise to a position-dependent light shift and hence to a state-dependent force~\cite{Leibfried2003, Lee2005}.
In the Lamb-Dicke approximation the Hamiltonian is

\begin{align}
H_\mathrm{LS}(t)=& \sum_{j=1,2}\sum_{s=\upa,\dna}-\frac{\hbar}{2} \left(e^{i\phi_{z,j}}\Omega_{s_j}\right)\ketbra{s_j}_j \nonumber \\ 
 & \bigg[ 2 \cos(\omega t-\phi_0)+ \nonumber \\ 
 &  \sum_{m=\mathrm{ip,oop}} 2\eta_{m,j}\xi_{m,j}(a_m+\ad_m)\sin(\omega t-\phi_0)\bigg]
\label{eqn:wobble_ld_full_ham}
\end{align}

\noindent with $\phi_{z,j}$ the phase of the travelling standing wave at the location of ion $j$, $\Omega_{s_j}$ the magnitude of the light shift on spin state $s=\upa,\dna$ of ion $j$, spin operators $\ketbra{s_j}_j$ which are $\ketbra{s_1}\otimes I$ for $j=1$ and $I \otimes \ketbra{s_2}$ for $j=2$, $m$ the axial in-phase (ip) or out-of-phase (oop) motional mode of the two-ion crystal, $\eta_{m,j}$ the corresponding Lamb-Dicke factor, $\xi_{m,j}$ the normal mode vector element, where $\xi_m=\left(\xi_{m,1},\xi_{m,2}\right)$, $\xi_{\mathrm{ip}}=\frac{1}{\sqrt{2}}(1,1)$ and $\xi_{\mathrm{oop}}=\frac{1}{\sqrt{2}}(1,-1)$, and $\phi_0$ the initial phase of the laser field given by the difference of the phases between the two beams.
$\phi_0$ is the LS gate equivalent of $\phi_{k,j}$ for the MS gate.
In the interaction picture, after the rotating wave approximation and ignoring the ``carrier-like" $\cos(\omega t-\phi_0)$ term, which can be suppressed via pulse-shaping, this simplifies to 

\begin{align}
H_\mathrm{LS}(t)= \sum_{s_1,s_2=\upa,\dna} &\frac{i\hbar}{2} \left(\eta_1\xi_1\Omega_{s_1}+\eta_2\xi_2e^{i\phi_z}\Omega_{s_2}\right)\ketbra{s_1s_2}\cdot  \nonumber\\
& \Big[ a e^{i(\delta_gt-\phi_0)}-\ad e^{-i(\delta_gt-\phi_0)} \Big] 
\label{eqn:LS_ham}
\end{align}

\noindent where $\phi_z=\phi_{z,2}-\phi_{z,1}$ and the gate detuning $\delta_g$ is the detuning of the frequency difference of the two gate laser beams $\omega$ from the motional mode frequency $\delta_g=\omega-\omega_z$. 
For $\eta_1\xi_1\Omega_{s_1} = -\eta_2\xi_2e^{i\phi_z}\Omega_{s_2}$ this corresponds to a \zg Hamiltonian, and is equivalent to eqn.\,\ref{eqn:MS_ham} apart from the rotated basis of the Pauli operator.

For identical ion species, both Hamiltonians displace one parity of their respective two-particle spin bases (\upa,\dna/+,$-$) in motional phase space, and drive them on a loop trajectory until they return to the origin after the gate time $t_g=\frac{2\pi}{\delta_g}$.
These spin components acquire a geometric phase corresponding to the area their trajectory enclosed in phase space.
The opposite-parity spin components remain at the origin and therefore acquire no phase.
If the differential phase between the even and odd parity spin components equals $\pi/2$, the two qubits are in a maximally entangled state at the end of the gate operation \cite{Lee2005}. 

For mixed species, the Rabi frequencies and Lamb-Dicke factors for the two ion species are in general different.
Therefore the final simplification of the Hamiltonian for both the LS and MS gate is not possible, and additional Pauli operators remain that also cause a displacement of the opposite-parity spin-states.
Therefore part of the acquired two-qubit phase instead becomes a global phase, no longer contributing to entanglement.
This requires for example a larger Rabi frequency and gate area as compensation, which can lead to larger errors from photon scattering, decoherence, out-of-Lamb-Dicke effects etc.
We can quantify this effect in the ``gate efficiency", which we define as $\zeta = \left( \Phi_{\mathrm{odd}} - \Phi_{\mathrm{even}} \right)/\left( \Phi_{\mathrm{odd}} + \Phi_{\mathrm{even}} \right)$, where the even-parity phase $\Phi_{\mathrm{even}}$ is the geometric phase acquired by the even-parity spin-states $\ket{s_is_i}+\ket{s_js_j}$ and $\Phi_{\mathrm{odd}}$ equivalently by $\ket{s_is_j}+\ket{s_js_i}$, with $s_{i,j}=\upa,\dna$ for the LS gate and $s_{i,j}=+,-$ for the MS gate.
More equally balanced Rabi frequencies lead to a more efficient gate and therefore smaller error sources.


\section{Phases}

For a two-qubit system, the internal state phase space can be spanned by the four basis vectors

\begin{align}
\hat{\theta}_1 &=\mathrm{diag}(1,1,-1,-1)
&& \hat{\theta}_2 =\mathrm{diag}(1,-1,1,-1) \\
\hat{\psi} &=\mathrm{diag}(1,-1,-1,1)
&& \hat{\varphi} =\mathrm{diag}(1,1,1,1) \nonumber
\end{align}

\noindent with eigenvalues $\theta_1,\theta_2,\psi,\varphi$ consisting of two single-qubit phases $\hat{\theta}_1=\psz\otimes I$ and $\hat{\theta}_2=I \otimes \psz$, a two-qubit phase $\hat{\psi}=\psz\otimes \psz$ and a global phase $\hat{\varphi}=I \otimes I$.
For an ideal two-qubit operation we want $\psi=\pi/2$ and $\theta_1=\theta_2=0$ while the global phase can take an arbitrary value \footnote{If there are more than 2 qubits the value of the global phase needs to be deterministic so it can be tracked, as the phase is no longer global with respect to the other qubits.}.
When the qubits can be addressed individually, which is trivial for mixed species, single-qubit phases can be corrected if they are deterministic.
This can be achieved by progressing the phase of the local oscillator of the driving field.
Additional two-qubit phases can be compensated by adjusting the Rabi frequency of the gate lasers.
Therefore, phases only negatively affect the fidelity if they are not calibrated correctly or vary between different shots of the experiment in an uncontrolled manner.

We can distinguish between the internal atomic phases, depending on the electronic qubit transition frequency \wo\ and the motional frequency \wz, as well as the external phases $\phi_0$, $\phi_z$ of the driving fields, i.e. lasers, microwaves and rf, at the respective positions of the ions.
The qubit phase $\phi_q$ is not visible in the Hamiltonian, as the Hamiltonian is in the rotating frame with respect to \wo, and therefore $\phi_q=0$ by definition.
Any changes in the qubit frequency mean the rotation velocity of the frame changes, and therefore a phase accumulates between the qubit and the fixed-frequency local oscillator of the lasers, rf and microwaves $\phi_0$.
This is equivalent to a change of $\phi_0$.
Unless otherwise specified, we use `phase' to refer to a single-qubit phase.
While the two qubits need not have a fixed phase relation with each other (for mixed species they precess at different frequencies), the phase relation between a qubit and its driving field is often important.
The relative phase of the two driving fields at each ion ($\phi_{z,j},\phi_{k,j}$) also matters, as it determines the phase of the force exciting the motion of the ions.
Here we distinguish four different time regimes:\\
(i) phase coherence over the duration of a single gate loop, taking $\sim50\us$, is always important;\\
(ii) phase coherence between different gates in a longer sequence of gates, such as randomised benchmarking or algorithms, on the timescale of $50\us-10$\,ms only matters for certain gate types and implementations;\\
(iii) phase coherence between different repetitions of the same gate sequence, over $10\ms-5\,\mathrm{min}$ can lead to a different final phase of the state and therefore loss of contrast during averaging;\\
(iv) phase coherence between different cycles of calibration, ranging from minutes to hours does not affect the measured fidelity but might matter in more large-scale systems.

The laser phase $\phi_0$ is randomised between each repetition of an experimental sequence, due to the limited phase coherence of the laser as well as path length fluctuations.
In addition, the different ion positions lead to a fixed difference of the laser phase of $\Delta\phi_z=\Delta k \Delta z$ at the two ion positions, where $\Delta k=|{\bm{k_1-k_2}}|$ is the wave vector difference of the gate beams.
While the laser phase can be actively stabilised~\cite{schmiegelow2016phase, saner2023breaking}, this adds experimental complexity and is not implemented in most experiments.
Changes in the internal phase of the qubit are predominantly caused by magnetic field fluctuations and light shifts for the electronic state, and noise and drifts in the trapping potential for the motional frequencies.

Slow drifts of these relative phases can be cancelled by employing first-order Walsh modulation~\cite{hayes2012coherent}.
For this, the gate is performed in a 2-loop configuration, where the gate is performed twice in sequence but with reduced power, so that in each loop only half of the total required two-qubit phase is acquired.
For the LS gate, embedding the gate in a spin echo sequence, rather than a Ramsey sequence, with the central $\pi$-pulse between the two gate pulses, cancels drifts of both the motional frequency and the qubit frequency.
For the MS gate, flipping the phase of the second loop by $\pi$ rad. cancels drifts of the motional frequencies, but not of the qubit frequency \wo\!.

Parameter mis-calibrations and offsets in different types of phases at the end of a single gate cause different types of errors that require different means for cancellation.
Imperfect closure of the phase-space loops, e.g. caused by a mismatch in the gate detuning and time can be cancelled by a two-loop Walsh modulation.
Acquisition of a global phase $\varphi$ caused by a mismatch between the acquired geometric phases of the different spin parities does not directly affect the fidelity of the gate but reduces its efficiency.
A mismatch of the total acquired two-qubit phase $\psi$, e.g. through too large or small laser powers, cannot be corrected, but is easy to calibrate.
This can be done by looking at the symmetry of the gate dynamics, i.e. the populations of the different spin parities when scanning the length of the gate pulse over several $t_\mathrm{g}$.
A mismatch between the single-qubit phases $\theta_1$ and/or $\theta_2$ still results in a perfectly entangled state.
For the LS gate and the phase-insensitive MS gate, as explained in the section on mixed-species MS gate implementations, single-qubit phases that change on timescales (ii)-(iv) do not cause any problems.
In the phase-sensitive MS gate, if the offset is known and constant, it can be corrected for by propagating the external phases accordingly.
Otherwise, however, acquisition of single-qubit phases causes significant errors once several gates are concatenated or a fixed-phase analysis pulse is applied.

\section{Mixed-species implementations}

Apart from different gate mechanisms, choices also have to be made about which qubit to do the gate operation on in each species (eg. optical, ground level hyperfine or Zeeman, or metastable qubit), and, for the LS gate, whether to use a common laser to address both species simultaneously or separate lasers for the different species.
Mixed species add several complicating factors compared to same-species gates:
First, there is usually no accessible magnetic field at which both species have a so-called clock transition that is first-order independent to magnetic field fluctuations.
While the conventional LS gate can only be performed on magnetic-field sensitive transitions \cite{Lee2005}, the MS gate is usually performed on clock transitions and will be affected by one qubit being sensitive to magnetic field noise for mixed species.
Secondly, depending on the transition frequencies, the gate lasers used for one species can cause significant light-shifts on the other species.
If the two gate operations are not sufficiently synchronised, these light shifts can vary over the course of the gate duration, and any intensity noise in the beam will translate into frequency noise for the light-shifted qubit.
Thirdly, the different mass of the ions can cause the crystal to tilt with respect to the trap axis for imperfect compensation of stray electric fields.
This can cause the gate lasers to couple to spectator modes -- here the radial modes of motion, which in our case are not cooled below the Doppler temperature and are therefore hotter than the axial modes, increasing errors from off-resonant excitation.
Fourthly, for strongly differing ion masses the reduced motional coupling also makes cooling less efficient and increases the required cooling duration or the final attainable temperature, especially for the radial modes.
Depending on the motional mode frequencies this can also increase errors stemming from Kerr cross-coupling \cite{Roos2008a}.
Since the local environment of the ions can vary, e.g. due to magnetic or electric field gradients across the trap, maintaining a fixed ion order is important when using mixed species.
Finally, the ion spacing, measured in multiples of the wavelength of the gate lasers, defines important phases such as $\phi_z$ in the LS gate and $\phi_{d,j}$ in the MS gate \cite{Lee2005}.
If different lasers are used for the two species this requires good relative phase stability between the two laser beams.

\begin{figure}
	\includegraphics[width=\linewidth]{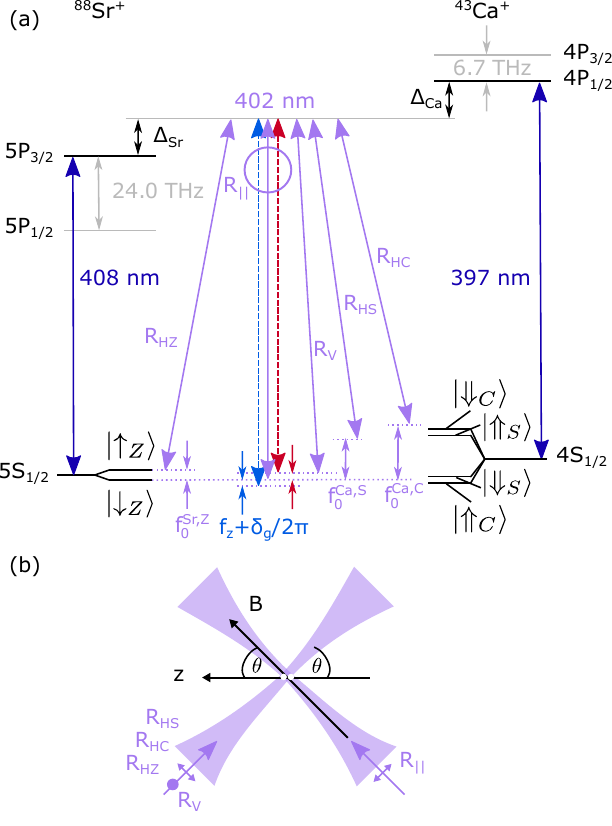}
	\caption{\textbf{ (a) Level structures} of \sr and \ct (not to scale), showing the strontium Zeeman (Z), calcium clock (C) and calcium stretch (S) qubits. At a static field $B=146\G$, these three qubits have frequencies $f_0^{\text{Sr,Z}} = 409$ MHz, $f_0^{\text{Ca,C}} = 3.200$ GHz, $f_0^{\text{Ca,S}} = 2.874$ GHz. The five gate beams ($\approx 402\nm$ wavelength) operate at Raman detunings $\Delta_\text{Sr} = +11.2$ THz and $\Delta_\text{Ca} = -9.0$ THz from the $5P_{3/2}$ and $4P_{1/2}$ levels in strontium and calcium respectively. $R_\parallel$ is shown at three separate frequencies, where the centre frequency is used for LS gates and the red and blue tones (dashed lines) form the bichromatic drive required for MS gates detuned by $\pm(f_z+\delta_g/2\pi)$.
    The notation $\ket{\Downarrow_C}$ is chosen for the higher energy clock qubit state, as the microwave pulse sequence prepares $\ket{\Downarrow_C}$ from $\ket{\Downarrow_S}$ and $\ket{\Uparrow_C}$ from $\ket{\Uparrow_S}$, and therefore the initial state for a MS gate operation is $\ket{\Downarrow_C \downarrow_Z}$.
    (b) Beam geometry and polarisation relative to the trap axis ($z$) and static field ($B$), where $\theta\approx45^\circ$.
    $R_\parallel$ is aligned parallel to the magnetic field, $R_V$ is polarised linear vertically with respect to the table, and $R_H$ horizontally, with frequencies addressing the clock ($R_{HC}$), stretch ($R_{HS}$) and Zeeman ($R_{HZ}$) qubits respectively. 
    } 
	\label{fig:beams}
\end{figure}

\subsection{Mixed-species LS gate}\label{sec:mixedspeciesLSgate}

\begin{figure}
\centering
\includegraphics[width=\linewidth]{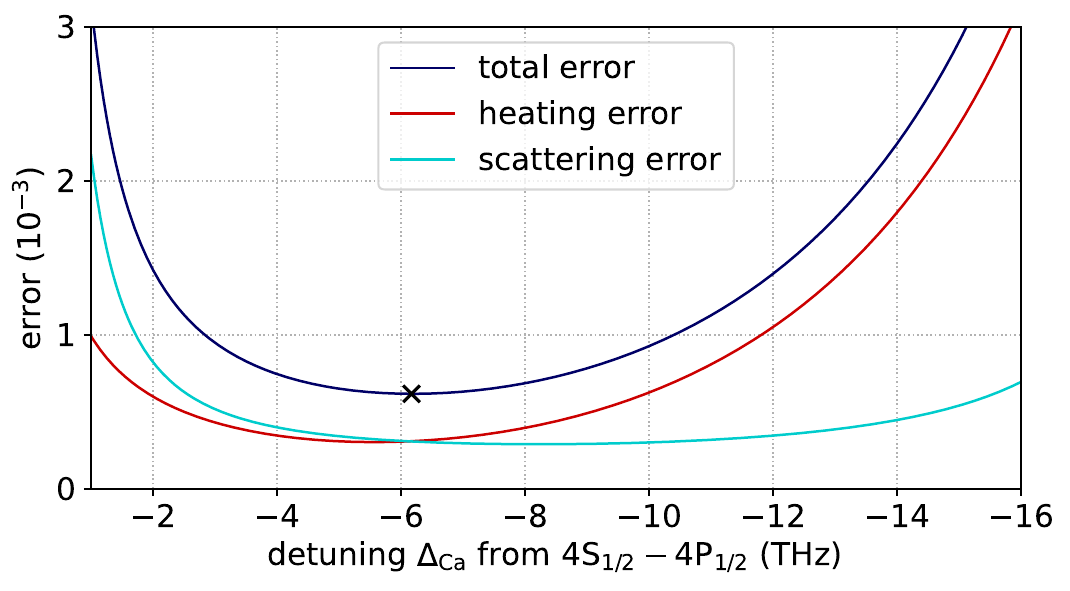}\\
\includegraphics[width=\linewidth]{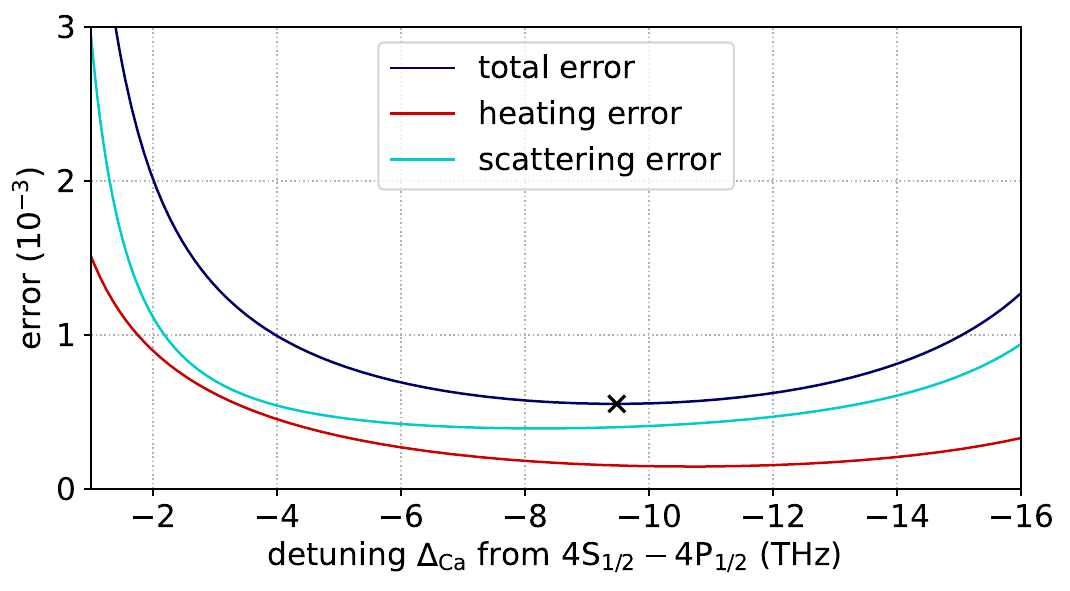}
\caption{{\bf Gate error vs Raman detuning:} Simulation of the error for a \ct$-$\sr\ entangling gate as a function of Raman detuning $\Delta_\mathrm{Ca}$.
Top: Gate on the axial ip mode ($1.55\MHz$), assuming a heating rate of $\dot{\bar{n}}=93\,\mathrm{quanta}/\mathrm{s}$. 
Bottom: Gate on the axial oop mode ($2.85\MHz$), assuming $\dot{\bar{n}}=27\,\mathrm{quanta}/\mathrm{s}$.
Both simulations assume a Raman beam radius ($1/e^2$ intensity) of $r_0=25\um$ at the ion and $P=70\mW$ in each beam. 
The black cross indicates the point of minimum total error.
The detuning used in the experiment was $\Delta_\mathrm{Ca}=-9.0\THz$.
The values for the heating rates are estimated from the measured heating rate of a single \ct ion.}
\label{fig:best_detuning_4388}
\end{figure}

The laser frequencies for the LS gate are independent of the qubit frequency.
Therefore a single laser can be used to drive a LS gate on different ion species, if the level structures share transitions that are sufficiently close in frequency \cite{Ballance2015}.
Due to the increased complexity and cost of requiring two separate gate lasers and the additional error sources associated with separate lasers outlined above, this is in general preferable.
For calcium and strontium sufficiently close transitions are available with the $20\THz$ separated $397\nm$ $4S_{1/2}-4P_{1/2}$ transition in calcium and the $408\nm$ $5S_{1/2}-5P_{3/2}$ transition in strontium, see Fig.\,\ref{fig:best_detuning_4388}.
Therefore a single pair of Raman laser-beams can be used to simultaneously drive a LS gate on the ground-level hyperfine/Zeeman qubits in these two species.
While a $\Delta \sim 10\THz$ Raman detuning requires larger laser powers to achieve a reasonable Rabi frequency (and therefore gate time), the large detuning also suppresses Raman scattering errors, which are often one of the largest error sources in trapped-ion laser gates \cite{Ozeri2005,Moore2023}.

The choice of the Raman detuning $\Delta$ affects the achievable gate fidelity in two different ways:
Firstly, via the Raman scattering errors from the two different species.
And secondly, via the gate efficiency due to the asymmetry in the Rabi frequencies.
A reduced gate efficiency increases the required area of the loops in phase space, and therefore increases gate errors stemming from imperfect closure of these loops, such as imprecisions in the gate time or detuning~\cite{Schafer2018, Ballance2016}.
It also leads to an increase of the gate duration for fixed available laser power.
This increases errors from heating and decoherence-type sources.
Combining these different error sources yields an optimal Raman detuning, see Figure\,\ref{fig:best_detuning_4388}.

The LS gate can only be implemented on qubits that are sensitive to magnetic-field changes, as for clock qubits the differential light shift that provides the spin-dependent force is cancelled to a high degree~\cite{Lee2005,langer2005long,Baldwin2021}.
This requires low magnetic field drifts and noise.
An advantage of the LS gate is its insensitivity to the long-term phase stability of the gate laser, as the gate operates in a basis orthogonal to single-qubit operations.
These can be performed using microwave or rf sources which tend to have considerably better phase coherence compared to lasers.
This advantage becomes especially important when longer sequences of gates are implemented, for example in algorithms or randomised benchmarking.


\subsection{Mixed-species MS gate}

As the driving fields for the MS gate are close to resonance with the qubit transition, which typically differs strongly for different ion species, different lasers have to be used for the two species.
The direct dependence on the qubit frequency also means that a stable transition frequency is of particular importance.
However, magnetic field-independent states usually occur at different magnitudes of {\bf B} for different species of ion (aside from $\bm{\mathrm{B}}= 0$), meaning that at least one species will be sensitive to magnetic field fluctuations.
Having to use different lasers for each species means that the gate can also be performed on different types of qubits in each ion, for example a ground level (Zeeman or hyperfine) qubit in one species and an optical qubit (between the S and D manifolds) in the other.
In the case of calcium--strontium this is not desirable though, as while the Raman polarisations can be optimised such that the differential light shift on the ground state vanishes, the absolute light shift on the ground state levels is still significant. 
Due to the proximity of the wavelengths the Raman lasers for one species therefore produce a significant light shift on the ground state of the optical qubit of the other species, shifting its transition frequency whenever they are on.
While this error can be suppressed by choosing a different Raman detuning and accounting for the light shift in the frequency of the laser driving the optical qubit, this approach is more complex and achieves lower fidelity; we therefore do not discussed it further.
Instead, we can again take advantage of the similar transition frequencies in calcium and strontium by using the bichromatic Raman beam jointly for both ion species, while using different frequencies for the second beam to address the distinct qubit frequencies, see Fig.\,\ref{fig:beams}.

\begin{figure}
\includegraphics[width=\linewidth]{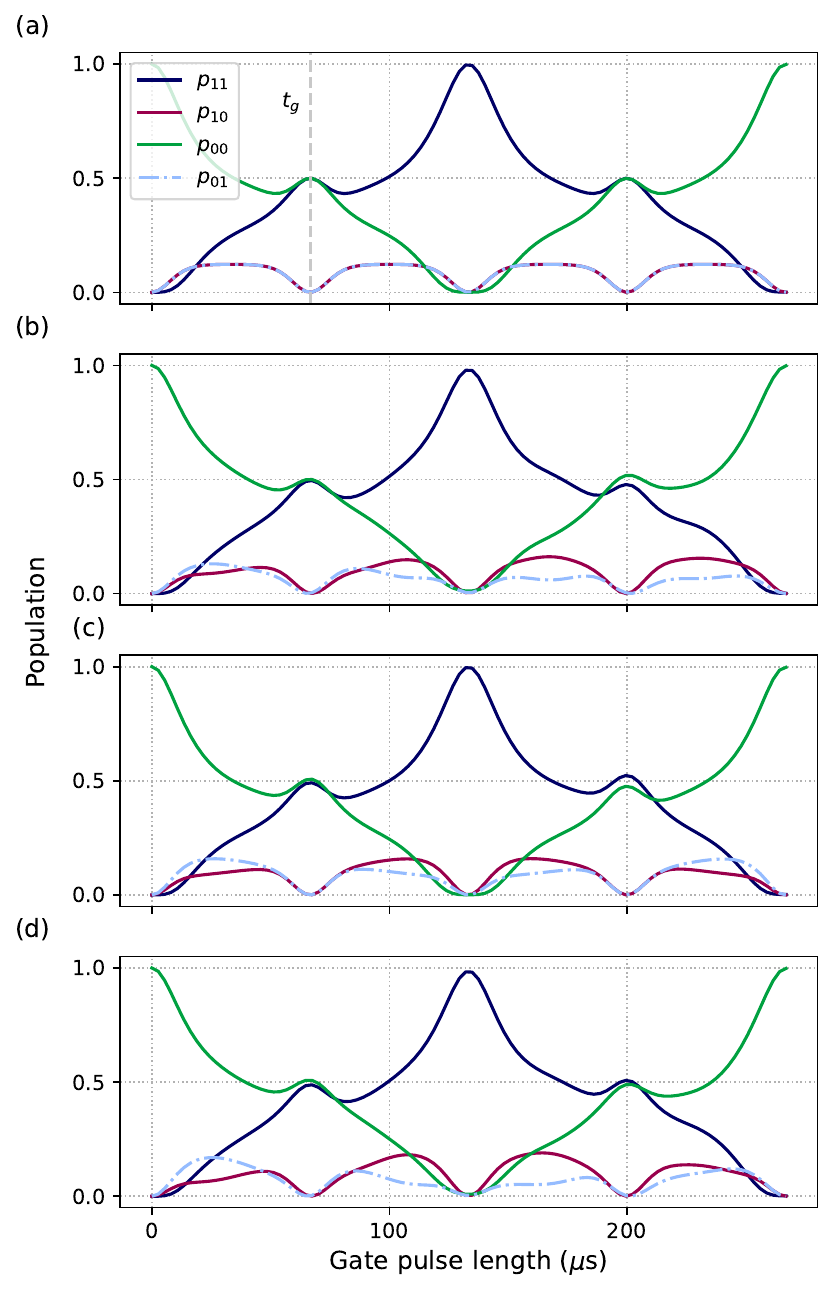}
\caption{\textbf{Mixed species gate dynamics}:
For mixed-species \MS gates, both unequal intensities of the bichromatic tones and different intensities on each ion species lead to asymmetries in the populations when scanning the length of the gate pulse.
The effect of these asymmetries can be distinguished by looking at all four different spin parities $p_{00}=\ket{\Downarrow\downarrow}$, $p_{01}=\ket{\Downarrow\uparrow}$, $p_{10}=\ket{\Uparrow\downarrow}$, $p_{11}=\ket{\Uparrow\uparrow}$.
\textbf{(a)} Simulation of ideal 2-loop gate dynamics. The ideal gate time is indicated through a dashed line.
\textbf{(b)} Gate dynamics with $10\%$ asymmetry in the intensity of the bichromatic tones.
Calibration of equally strong sidebands is best performed using single-species gates to reduce the degrees of freedom. 
\textbf{(c)} Gate dynamics with $10\%$ asymmetry in the intensity for each species. 
\textbf{(d)} Gate dynamics with $10\%$ asymmetry in the intensity of both the bichromatic tones and for each ion species. }
\label{fig:dynamics}
\end{figure}

\begin{figure*}
\centering
\includegraphics[width=\textwidth]{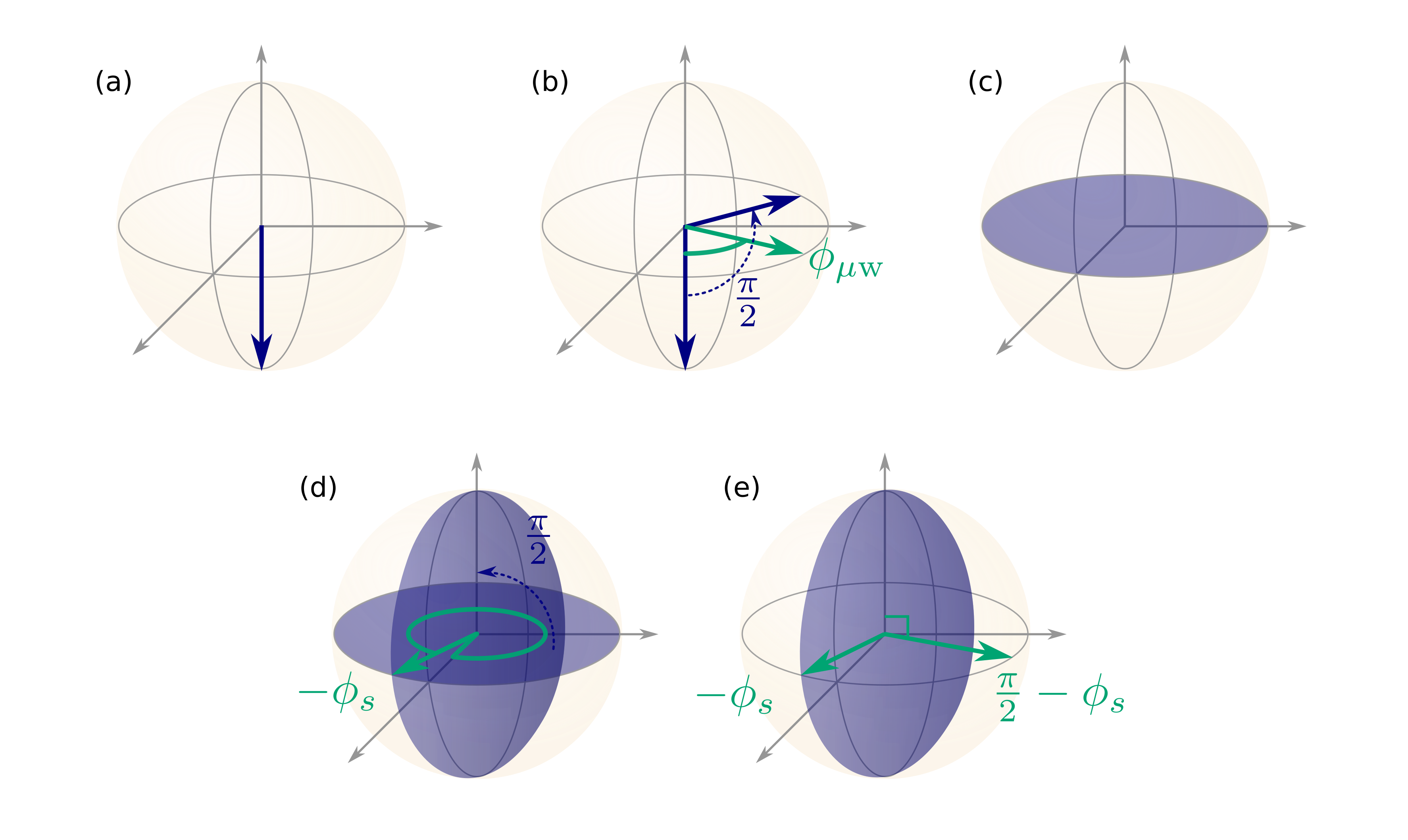}
\caption{\textbf{Phase insensitive MS sequence}: \textbf{(a)} The ions are prepared in \dn and \Dn. 
\textbf{(b)} A microwave (rf) $\pi/2$ pulse is applied that transfers the state into the $xy$-plane. 
The operator is a $\sigma_{-\phi}$ operator with $\phi=\phi_{\mu\mathrm{w}} (\phi_{\mathrm{rf}})$. 
There is no fixed phase relationship between $\phi_{\mu\mathrm{w}}$ and $\phi_{\mathrm{rf}}$. 
\textbf{(c)} As the phase $\phi_{\mu\mathrm{w}} (\phi_{\mathrm{rf}})$ is arbitrary with respect to the laser phase $\phi_s$, the qubit state is now at an arbitrary angle in the $xy$-plane of the Bloch sphere from the perspective of the gate lasers.
\textbf{(d)} We now apply a $\pi/2$ carrier pulse with the gate laser. 
The operator is a $\sigma_{-\phi}$ operator with $\phi=\phi_s$. This moves the qubit state into a plane orthogonal to $xy$ and containing the vector $\phi_s$. 
Again, $\phi_s$ for the different ion species have no fixed phase relationship, as they are derived from different lasers. 
Both $\phi_s$ also have no fixed phase relationship to $\phi_{\mu\mathrm{w}} (\phi_{\mathrm{rf}})$.
\textbf{(e)} The gate operator is $\sigma_{\pi/2-\phi_s}$ and therefore orthogonal to the qubit's state. 
This, therefore, leads to a maximally entangling gate independent of $\phi_s$. 
If no $\pi/2$ pulses are applied, the initial state is \ddKet, which is also orthogonal to  $\sigma_{\pi/2-\phi_s}$ for all $\phi_s$. 
The gate operation leads to the acquisition of geometric phases and therefore entanglement between the two qubits, while leaving the state on the single-ion Bloch sphere at end of the gate unchanged.
Therefore, applying another laser carrier pulse with fixed phase $\pi-\phi_s$, followed by a microwave (rf) $\pi/2$ pulse of phase $\phi_{\mu\mathrm{w}} (\phi_{\mathrm{rf}})$ returns the state back onto the \zh -axis.
The initial and final laser $\pi/2$ pulses write and unwrite the phase of the laser onto the qubit, so that only fluctuations during the gate matter.
}
\label{fig:phase_ins}
\end{figure*}

The \MS gate can be implemented in two different configurations \cite{Lee2005, langer2006high}: (i) the phase-sensitive configuration, where $\delta_k=\pm(\wz+\dlg)+ \wo$ and $\phi_d\approx 0$ and (ii) the phase-insensitive configuration, where $\delta_k=\wz+\dlg\pm \wo$ and $\phi_s\approx 0$.
For hyperfine qubits, the phase-insensitive configuration is experimentally less convenient, as then the $\sim\GHz$ hyperfine splitting has to be bridged twice in opposite directions.
For mixed species, this configuration has an additional disadvantage:
because $\phi_d \neq 0$ the efficiency of the gate is reduced if $\phi_d$, which determines the phase of the force on each ion, is not equal for both ions.
As $\phi_d$ depends on both the ion position and the laser phase, this can be achieved by adjusting the ion spacing for a common gate beam.
If there are different laser beams driving the gate for each species, this trick no longer works and the absolute phase of both lasers would have to be stabilised.
In the phase-sensitive configuration this problem does not occur, since $\phi_d\approx 0$.
Here instead $\phi_s\neq 0$, which determines the basis $\sigma_{\frac{\pi}{2}-\phi_{s,j},j}$ in which the gate is performed.
As long as all operations, i.e. the gate and any single-qubit rotations, are performed by the same laser this does not cause any difficulties, as $\phi_s$ similarly defines the basis of the single-qubit operations.
However, for longer gate sequences such as algorithms or randomised benchmarking this fixed dependence on the laser phase will degrade the fidelity due to decoherence of the laser phase.
By applying additional laser and microwave $\pi/2$-pulses \cite{Lee2005,Tan2015}, the gate can be made phase-insensitive, see Fig.\,\ref{fig:phase_ins}.
Here the phase-coherence requirement is transferred from the laser to the microwave local oscillator, which usually has a significantly longer coherence time.

Due to the larger number of lasers and frequency tones, calibration of the gate intensities is more complex for mixed-species \MS gates.
By analysing the asymmetries arising in scans of the gate pulse length and comparing them to theoretical simulations, mis-set parameters can be identified and corrected.
The simulations were performed using qutip~\cite{johansson2012qutip,johansson2023qutip}, and the code is available on GitHub \footnote{https://github.com/OxfordIonTrapGroup/gate-simulations}.


\section{Experimental results}

We implement mixed-species LS gates and MS gates in the same experimental system, a macroscopic `blade' ion trap described in detail in \cite{Thirumalai2019}.
The gates are performed on a two-qubit, mixed-species crystal of \ct and \sr in an external magnetic field {\bf B} = 146\,G.
This field strength gives access to a magnetic-field-insensitive ``clock" qubit in \ct, formed between two hyperfine states $\{\ket{\Uparrow_C}=\ket{F=4, m_F=0}, \ket{\Downarrow_C}=\ket{F=3, m_F=+1}\}$ in the ground level.
The states $\{\ket{\Uparrow_S}=\ket{F=3, m_F=+3}, \ket{\Downarrow_S}=\ket{F=4, m_F=+4}\}$ form the field-sensitive ``stretch" qubit.
In \sr, the Zeeman qubit $\{\ket{\uparrow_Z}, \ket{\downarrow_Z}\}$ is formed by the two states in the ground level.
These qubits may be driven directly using microwave (MW) or radiofrequency (RF) fields at $\approx$ 3 GHz in calcium and 409 MHz in strontium, or via Raman transitions coupling them to the P levels.

The gate lasers are two phase-locked, frequency-doubled Ti:sapphire lasers, operated at a wavelength of $\approx 402\nm$, such that we may drive S $\leftrightarrow$ P Raman transitions in both \ct and \sr simultaneously.
The secondary laser is locked at a frequency 1.6 GHz below the master laser, in order to span the \ct hyperfine splitting after frequency doubling. From the secondary laser we derive two horizontally polarised beams, referred to as $R_{HS}$ and $R_{HC}$, and from the master we derive one horizontally polarised beam, $R_{HZ}$, one vertically polarised beam, $R_V$, and one beam, $R_\parallel$, which enters the vacuum system parallel to the direction of the quantisation field, see Figure \ref{fig:beams}b, and is therefore $\sigma^\pm$-polarised.
In this geometry, different beams pairs allow us to drive the clock and stretch qubits in \ct and the Zeeman qubit in \sr in configurations that are sensitive to the ions' axial motion, as required for the MS gate, and to generate the travelling standing wave required for the LS gate; the various combinations are outlined in Table \ref{tab:ramanbeampairs}.
Their frequency offsets and polarisations are shown in Figure \ref{fig:beams}.
The axial mode frequencies and Lamb-Dicke parameters are listed in Table \ref{tab:ldparams}.

The Raman detuning $\Delta_\text{Ca}=-9.0\THz$ is chosen to approximately minimise the total error contribution from photon scattering and from motional mode heating, as shown in Figure \ref{fig:best_detuning_4388}.

\begin{table}[h!]
	\begin{center}
		\def\arraystretch{1.2}
		\begin{tabular}{ c|c }
			beam pair & drives \\
			\hline
			$R_\parallel + R_{HS}$ & \ct stretch (MS) \\
			$R_\parallel + R_{HC}$ & \ct clock (MS) \\
			$R_\parallel + R_{HZ}$ & \sr Zeeman (MS) \\
			$R_\parallel + R_V$ & LS gate \\
		\end{tabular}
		\vspace{0.1in}
		\caption{Raman beam combinations for entangling gates.}
		\label{tab:ramanbeampairs}
	\end{center}
\end{table}
\begin{table}[h!]
	\begin{center}
		\def\arraystretch{1.5}
		\begin{tabular}{ c|c c c }
			mode & frequency $f_z$ (MHz) & $\eta_{\text{Ca}}$ & $\eta_{\text{Sr}}$ \\
			\hline
			in-phase & 1.49 & 0.090 & 0.124 \\
			out-of-phase & 2.91 & 0.127 & $-$0.045 \\
		\end{tabular}
		\vspace{0.1in}
		\caption{Axial motional mode frequencies for \ca and \sr, and the corresponding Lamb--Dicke parameter $\eta$ for each ion in the Raman beams.}
		\label{tab:ldparams}
	\end{center}
\end{table} 

\subsection{LS gate}

We perform a LS gate on the stretch qubit in \ct and the Zeeman qubit in \sr.
The required frequency difference between the two Raman beams, $f_\text{z,oop} + \delta_g/2\pi$,  is independent of the two qubit frequencies and lies close to the out-of-phase motional mode frequency.
Therefore we may drive both species using the same beam pair, $R_V$ and $R_\parallel$, with both beams derived from the same laser and the frequency splitting  provided using an AOM.
For a maximally efficient gate, we set the distance between the ions to a half-integer multiple (3.57\,$\mu$m) of the effective wavelength $\lambda_\text{eff} = 402\,\text{nm}/\sqrt{2}$ of the standing wave. This results in maximal motional excitation on the even-parity spin states for a gate on the out-of-phase motional mode. The gate detuning $\delta_g = 2\pi \times -40\kHz$, and the power in each of the two beams is $60\mW$. 

The LS gate is implemented in a two-loop sequence, where the gate lasers are pulsed twice, for a total duration of $2 \times 2\pi/\delta_g=49.2\us$. Between the two laser pulses is a single-qubit spin-echo $\pi_x$-rotation on each qubit; for mixed-species gates this ensures that the total geometric phase acquired by the two even-parity spin states is the same, despite differing Rabi frequencies of the gate lasers on the two species. It also reduces sensitivity to correlated magnetic field noise over the gate duration (in the case of a constant offset, completely cancelling its effect on the acquired phase), since the LS gate operator commutes with $\sigma_z$ rotations.
The phase of the spin-dependent force during the second pulse is shifted by $\pi$ rad relative to that of the first, implementing a first-order Walsh modulation \cite{Hayes2012} which reduces sensitivity to imperfect closure of loops in phase space.
This is achieved by the combination of the spin-echo pulse and a shift in the laser phase by $-\delta_g t_\text{delay}$, where $t_\text{delay}$ is the time between the leading edges of the two pulses. 

Single-qubit pulses, including the spin-echo pulses, are implemented by simultaneous application of MW and RF signals at the qubit frequencies, 2.874\,GHz and 409\,MHz, to one of the blade-shaped trapping electrodes and to an in-vacuum antenna respectively.

Starting from the state $\ket{\Downarrow_S \downarrow_Z}$, we generate the maximally-entangled state $(1/\sqrt{2})(\ket{\Uparrow_S\uparrow_Z} - \mathrm{i}\ket{\Downarrow_S\downarrow_Z})$ by first applying single-qubit $\pi/2$ $x$-rotations to each ion, followed by the LS gate and a final pair of single-qubit $\pi/2$ $x$-rotations.

\subsection{MS gate}

MS gates are driven in the phase-sensitive configuration. The $R_\parallel$ beam is split into two frequency components at $\pm(f_\text{z,ip} + \delta_g/2\pi)$ from its base frequency. As the Raman frequency splitting required for an MS gate depends on the qubit frequency, the bichromatic beam must be used in conjunction with two other beams, one for each species. For strontium, we use $R_{HZ}$, with a frequency 409\,MHz lower than the centre frequency of $R_\parallel$. For calcium, we use $R_{HS}$ for the stretch qubit or $R_{HC}$ for the clock qubit, at 2.874\,GHz and 3.200\,GHz below $R_\parallel$ respectively. The power in each of the two bichromatic tones in $R_\parallel$ is 120\,mW. We set the other powers to achieve the same Rabi frequency on the blue sideband of the axial in-phase mode on each species ($R_{HS}$ = 17\,mW, $R_{HC}$ = 18\,mW, $R_{HZ}$ = 8\,mW)\footnote{Note the spot sizes of the Raman beams were different between the LS gate implementation and the MS gate, so their powers should not be compared across implementations.}.
The gate detuning is $\delta_g = 2\pi \times 40$\,kHz, and the total gate time is $2 \times 2\pi/\delta_g=50\us$ for a two-loop gate.

Starting from $\ket{\Downarrow_S \downarrow_Z}$, the ideal MS gate generates the maximally-entangled state  $(1/\sqrt{2})(\ket{\Uparrow_S\uparrow_Z} - \mathrm{i} e^{i(\phi_{s,1} + \phi_{s,2})}\ket{\Downarrow_S\downarrow_Z})$.
The gate is again implemented in a first-order Walsh sequence, where the phases of the two $R_\parallel$ tones are adjusted by $\pi \pm \delta_g t_\text{delay}$ for the second pulse relative to the first.

Any additional single-qubit rotations are implemented via the same lasers as are used for the gate drive.

\subsection{Results}

We characterise the performance of each gate using partial state tomography of a maximally entangled Bell state \cite{Leibfried2003}, considering the total populations in $\ket{\Uparrow\uparrow}$ and $\ket{\Downarrow\downarrow}$ after the gate operation, and the parity contrast observed when applying additional single-qubit $\pi/2$-rotations. The parity contrast is measured using a two-point method where the additional rotations are performed about axes $\phi_\frac{\pi}{2} = \phi'$ and $\phi_\frac{\pi}{2} = \phi' + \pi/2$  in the $x$-$y$ plane, and $\phi'$ is the phase for which the contrast should be maximised for the expected Bell state. This two-point measurement ensures that the inferred fidelity is sensitive to changes of the phase of the Bell state we create, which will show up as a loss of inferred parity contrast.

\begin{figure}
    \includegraphics[width=\linewidth]{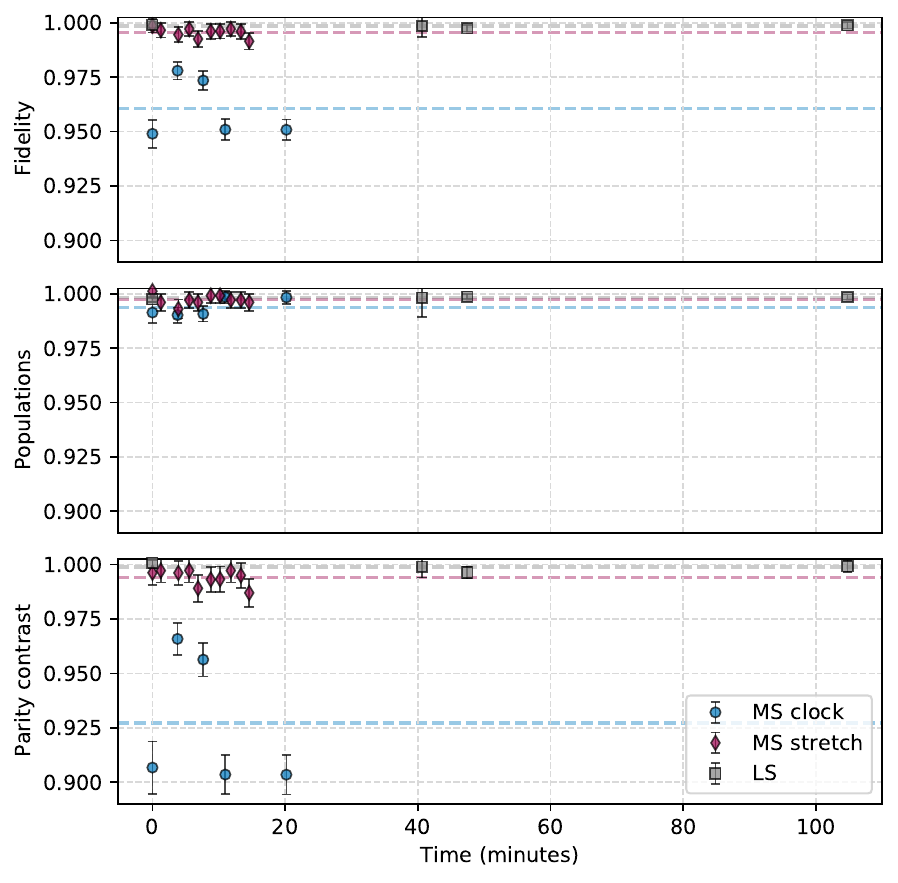}
    \caption{\textbf{Fidelity measurements:} Partial state tomography of maximally-entangled Bell states generated by $\mathrm{Ca}^+-\mathrm{Sr}^+$ mixed-species entangling gates. The fidelity with which we generate the desired Bell state is given by $F = \frac{1}{2}(\text{populations} + \text{parity contrast})$. The average fidelities, populations and parity contrasts are indicated with dashed lines for each gate type. The average fidelities are 96(1)\% for the clock--Zeeman MS gate, 99.6(2)\% for the stretch--Zeeman MS gate, and 99.8(1)\% for the LS gate. Average populations are above 99.4\% for all gate types, but the poorer and fluctuating parity contrast of the clock--Zeeman MS gate reduces the fidelity for this gate.}
    \label{fig:pst}
\end{figure}

We measure a Bell-state fidelity of 99.8(1)\% for the LS gate, 99.6(2)\% for the stretch--Zeeman MS gate and 96(1)\% for the clock--Zeeman MS gate.
Each fidelity is corrected for independently-measured state-preparation and readout errors for the relevant qubit states, the average of which is 0.32(8)\%.
Results are shown in Figure \ref{fig:pst}.
More detailed experimental results for the LS gate, including a consistent fidelity measurement by interleaved randomised benchmarking, can be found in \cite{hughes2020benchmarking}. 

\begin{figure}
    \includegraphics[width=\linewidth]{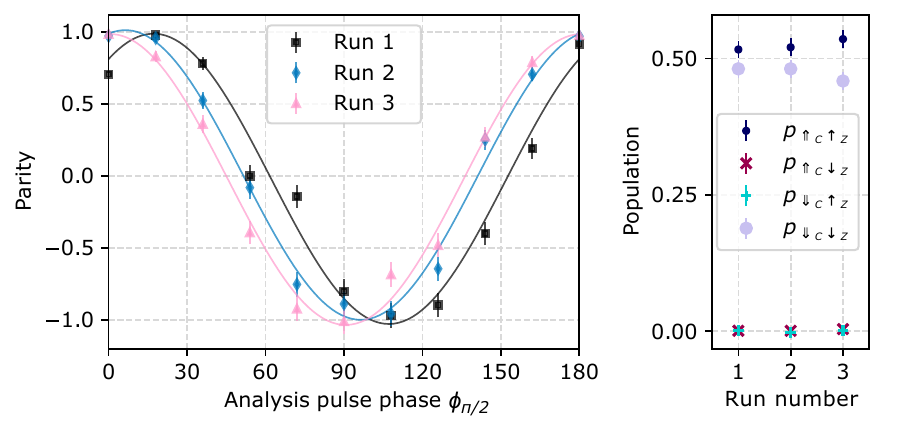}
    \caption{\textbf{Drift of parity phase}: Three parity contrast measurements for the clock--Zeeman MS gate, taken a few minutes apart. 11 different values of the analysis $\pi/2$-pulse phase $\phi_\frac{\pi}{2}$ are used, rather than the two-point method used for the Bell state fidelity measurement, and we measure the parity $P = p_{\Uparrow\uparrow} + p_{\Downarrow\downarrow} - (p_{\Uparrow\downarrow} + p_{\Downarrow\uparrow})$ of the final state. A least-squares sinusoidal fit of the form $P = P_0 + C\text{sin}(2(\phi_\frac{\pi}{2} - \phi_p))$, where $C$ is the parity contrast, illustrates the drift in the phase offset $\phi_p$ of the parity oscillation.}
    \label{fig:parityscans}
\end{figure}

It can be seen that the significantly lower Bell-state fidelity for the clock MS gate is due mostly to an apparent loss of parity contrast. Upon closer inspection, the contrast of parity oscillations remains approximately constant over time but the phase offset changes (see Figure \ref{fig:parityscans}), indicating a change in the phase of the prepared Bell state and leading to a lower inferred contrast when using the two-point measurement. This is attributable to slow drifts in the external magnetic field strength resulting in qubit frequency drifts of the magnetic-field-sensitive strontium qubit. The LS gate, acting in an orthogonal basis and allowing for the use of a spin-echo pulse, is insensitive to this drift. One would expect the stretch--Zeeman MS gate to be susceptible to the same error source, yet we measure a Bell-state fidelity comparable to the LS gate. This is because for the input state $\ket{\Downarrow\downarrow}$ as used in this work, the Bell-state phase offset depends on the sum of the qubit frequency offsets $\delta_0^\text{Ca} + \delta_0^\text{Sr}$. In this particular case, the magnetic-field sensitivities of the \sr Zeeman qubit (2.80\,MHz/G) and the \ct stretch qubit ($-2.36\MHz/\mathrm{G}$) are almost equal and opposite, meaning that $\delta_0^\text{Ca} + \delta_0^\text{Sr}$ remains small and the effect of magnetic field drift is suppressed. The effect is most pronounced in the clock--Zeeman MS gate, where the sensitivity of the clock qubit is negligible compared to that of the Zeeman qubit. Note, however, that for an input state $\ket{\Downarrow\uparrow}$  or $\ket{\Uparrow\downarrow}$ this error would depend instead on the \textit{difference} in qubit frequency offsets between the two species and the stretch--Zeeman MS gate would perform more poorly, highlighting the need for averaging over input states to characterise the fidelity of the gate operation itself -- for example using randomised benchmarking. To run such a protocol involving longer sequences of MS gates without incurring errors due to variation of this phase would require improved stabilisation of the external magnetic field, shielding of the qubits from fluctuations, further dynamical decoupling techniques, or tracking and feedforward onto the MS gate spin phase, $\phi_s$.


\begin{figure}
	\includegraphics[width=\linewidth]{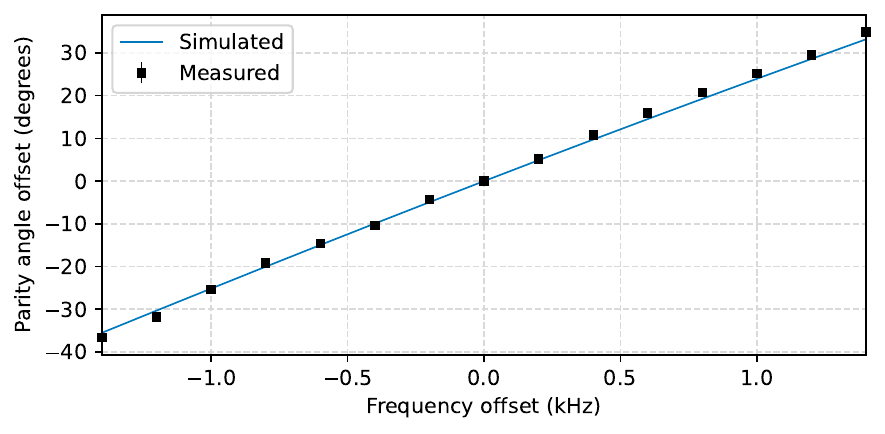}
	\caption{\textbf{Parity phase dependence on detuning from qubit frequency}: Experimentally-measured offset $\phi_p$ in parity oscillations when implementing a \ct$-$\ct MS gate on the clock qubit. We intentionally detune one of the gate lasers from its ideal frequency to replicate a qubit frequency offset. The results match well with simulation.}
	\label{fig:bfieldparity}
\end{figure}

To demonstrate the effect of magnetic field drift on the Bell-state phase, we perform a (same-species) MS gate on a \ct--\ct crystal using $R_\parallel$ and $R_{HC}$. We drive the gate on the clock qubit and add an offset to the frequency of the $R_{HC}$ laser, to model a qubit frequency offset of the typical size that we observe in \sr or on the \ct stretch qubit. These known shifts are typically $\pm 1$ kHz, corresponding to a change of $\pm 0.4$ mG in the external magnetic field strength as the field-generating coils vary in temperature. Driving the clock qubit allows us to introduce a controlled effective qubit frequency shift whilst remaining insensitive to the actual magnetic field drifts in the system. We measure the parity at several different values of $\phi_\frac{\pi}{2}$ in order to fit the phase offset in the parity fringe oscillations. The result is shown in Figure \ref{fig:bfieldparity} and agrees well with simulations.

\section{Conclusion}


In summary, we have implemented mixed-species LS and MS gates on the \sr Zeeman qubit and the \ct stretch qubit, achieving high and statistically consistent Bell-state fidelities as measured by partial state tomography.
We have also implemented a mixed-species MS gate using the \ct clock qubit, with a Bell-state fidelity limited by slow drifts in the external magnetic field.

The LS gate mechanism is less experimentally complex because we can use two beams from a single laser, single-qubit operations may straightforwardly be implemented with MW/RF drive, and we can protect against qubit frequency errors with a single spin-echo sequence.
The MS gate can be used directly on the clock qubit but requires extra laser frequencies and is less inherently robust to magnetic field drifts. To improve its performance, dynamical decoupling techniques could be used to protect against qubit frequency errors \cite{Harty2016}. In the configuration used in this work, the basis of the MS gate is determined by the optical phases of the gate lasers and is therefore sensitive to small optical path length fluctuations. For this reason, single-qubit pulses following MS gates were implemented using the same lasers as were used for the gate; however, as discussed in Figure \ref{fig:phase_ins}, additional $\pi/2$-rotations could be employed to remove this phase dependence and allow MW/RF fields to be used for general single-qubit operations as for the LS gate. 

Introducing additional degrees of freedom, like multiple driving tones, has been shown to make two-qubit entangling gates in same-species ion crystals more resilient against errors such as heating and drifts of motional frequencies~\cite{shapira2018robust,webb2018resilient}.
These techniques could be especially interesting for mixed-species gates, where the reduced gate efficiency leads to larger excursions of the states in motional phase space and exacerbates these types of errors.
While we only discuss laser-based gates here, magnetic-gradient-based gates have been shown to be an interesting alternative for high-fidelity gate operations \cite{Harty2016, srinivas2021high, loschnauer2024scalable}, where the better phase coherence of microwave and radiofrequency sources provides a useful advantage. To date, no mixed-species entangling gates have been implemented using laser-free methods, although an extension of the method in \cite{srinivas2021high} to mixed species is proposed in \cite{knaackthesis}, requiring the addition of two additional drive frequencies close to the qubit frequency of the second species. If entangling gates are to be implemented on motional modes with strongly asymmetric participation of each ion in the motion -- as opposed to the relatively mild asymmetry for the axial modes used in this work -- the gate speed will be limited; possible mitigations for this are also proposed in \cite{knaackthesis}.

An alternative to using mixed ion species is to use different types of qubits within the same ion, e.g. metastable and ground-state qubits~\cite{allcock2021omg, sotirova2024trapped, wang2025experimental}.
This has the advantage of providing spectral separation without an asymmetry in the ions' masses, yielding better motional coupling and allowing for easier shuttling, splitting and merging operations during transport.
Most considerations for mixed species in this paper apply similarly for mixed-qubit-type gates.  

\section{Acknowledgements}
V.~M.~S.\ was partly funded by Christ Church, Oxford, K.~T.\ by the U.~K.\ Defence Science and Technology Laboratory. C.~J.~B.\ was supported by a UKRI Future Leaders Fellowship and is a Director of Oxford Ionics Ltd. This work was funded by the U.K.\ EPSRC ``Quantum Computing and Simulation'' Hub (EP/T001011/1) and by the E.~U.\ Quantum Technology Flagship project, AQTION (Grant No. 820495).

\bibliography{bibliography}

\end{document}